\documentstyle{article}
\def\be{\begin{equation}}
\def\ee{\end{equation}}
\def\beq{\begin{equation}}
\def\eeq{\end{equation}}
\def\beqar{\begin{eqnarray}}
\def\eeqar{\end{eqnarray}}
\def\barr{\begin{array}}
\def\earr{\end{array}}

\def\and{\qquad {\rm and } \qquad}

\def\slp{p \hspace{-1ex}/}

\def\slp{p \hspace{-1ex}/}
\def\sls{s \hspace{-1ex}/}

\def\sp{\vec{s}_+}
\def\sm{\vec{s}_-}
\def\hp{h_+}
\def\hm{h_-}
\def\Kv{\vec{K}}
\def\pv{\vec{p}}
\def\g5{\gamma_5}
\newcommand{\nc}{\newcommand}
\nc{\jc}{\frac{1}{4}}  \nc{\sll}{S_{LL}}     \nc{\slr}{S_{LR}}
\nc{\srl}{S_{RL}}      \nc{\srr}{S_{RR}}     \nc{\vll}{V_{LL}}
\nc{\vlr}{V_{LR}}      \nc{\vrl}{V_{RL}}     \nc{\vrr}{V_{RR}}
\nc{\tll}{T_{LL}}      \nc{\tlrs}{T_{LR}}    \nc{\trl}{T_{RL}}
\nc{\trr}{T_{RR}}      \nc{\slld}{S_{LL}^D}  \nc{\slrd}{S_{LR}^D}
\nc{\srld}{S_{RL}^D}   \nc{\srrd}{S_{RR}^D}  \nc{\vlld}{V_{LL}^D}
\nc{\vlrd}{V_{LR}^D}   \nc{\vrld}{V_{RL}^D}  \nc{\vrrd}{V_{RR}^D}
\nc{\tlld}{T_{LL}^D}   \nc{\tlrd}{T_{LR}^D}  \nc{\trld}{T_{RL}^D}
\nc{\trrd}{T_{RR}^D}   \nc{\aqde}{\alpha_{qde}}
\nc{\alq}{\alpha_{\ell q}}        \nc{\alqp}{\alpha_{\ell q'}}
\nc{\alqt}{\alpha_{\ell q}^{(3)}} \nc{\alqtc}{\alpha_{\ell
q}^{(3)*}} \nc{\alqj}{\alpha_{\ell q}^{(1)}}
\nc{\alqjc}{\alpha_{\ell q}^{(1)*}} \nc{\aeu}{\alpha_{eu}}
\nc{\alu}{\alpha_{\ell u}} \nc{\aqe}{\alpha_{qe}}
\nc{\ber}{\begin{eqnarray*}} \nc{\enr}{\end{eqnarray*}}
\nc{\jmpb}{(1-\beta)/(1+\beta)} \nc{\wspR}{r}      \nc{\varx}{x}
\nc{\bt}{\beta}

\nc{\non}{\nonumber} \nc{\lspace}{\;\;\;\;\;\;\;\;\;\;}
\nc{\llspace}{\lspace \lspace}
\nc{\jnl}{\frac{1}{{\mit\Lambda}^2}} \nc{\jd}{\frac{1}{2}}
\nc{\comment}[1]{}

\begin{document}
\begin{flushright}
 hep-ph/0607013
\end{flushright}
\begin{center}
{\bf {\large \bf Probing space-time structure of new physics with
polarized beams at the ILC}}\footnote{Talk given at LCWS06,
Linear Collider Workshop, 9-13 March 2006, Bangalore, India}\\
\medskip
{\bf B. Ananthanarayan}\\    
\medskip
{\it Centre for High Energy Physics, Indian Institute of Science,\\ 
Bangalore 560 012, India}\\
\end{center}
\begin{abstract}
{At the International Linear Collider 
large beam polarization of both the electron and positron beams
will enhance the signature of physics due to interactions that are
beyond the Standard Model. 
Here we review our recently obtained results on
a general model independent method of determining 
for an arbitary one-particle inclusive state
the space-time structure of such new physics through
the beam polarization dependence and angular distribution
of the final state particle.}
\end{abstract}

\section{Introduction}
At the International Linear Collider, the possibility of
considerable beam polarization has led to a series of
investigations on using 
this as a diagnostic aid for new physics arising
due to Beyond the Standard Model interactions
(BSM) (for a recent review, see ref.~\cite{hep-ph/0507011}).
We recently considered the possibility of observing CP violating
asymmetries in $t\overline{t}$ production with transversely
polarized beam~\cite{hep-ph/0309260}.  It was shown that only interactions 
that transform under the Lorentz transformations as pseudo-scalar ($P$),
scalar ($S$)
and tensor ($T$) interactions could contribute.  
The fact that axial-vector ($A$) and vector ($V$) interactions will
not contribute to CP violating asymmetries with transverse beam
polarization could have partly been deduced from some general results
available in the literature for a general single-particle
inclusive process, albeit for the case that the 'new physics'
amplitudes interfere with the QED part of the standard model 
(SM) amplitudes~\cite{DR}.

These general results do not directly apply to other processes of 
interest, e.g.,
$Z \gamma$ production, where the SM production goes via t- and u- channel
amplitudes, in contrast to the case of $t\overline{t}$ production where
the SM production goes via s- channel amplitudes.  Therefore, the
general results available in the literature do not apply to the
latter process and a case by case study has had to be 
performed~\cite{hep-ph/0404106,hep-ph/0410084,hep-ph/0507037}.

Indeed, as mentioned earlier, general results involving both the
QED as well as the neutral current amplitudes for single-particle
inclusive process would be of general interest, as well as
the extension to t- and u-channel processes.  Here we review the
our recently obtained results on the subject~\cite{hep-ph/0601199}.
It may also be noted that our results are sufficiently general to
permit a discussion of features of, e.g., chargino and neutralino
production in the minimal supersymmetric standard model (MSSM).

We note here that we do not give an extended bibliography on
the subject and instead refer to the same in ref.~\cite{hep-ph/0601199}.

\section{Correlations and their features}
The process of interest to us here is the one-particle inclusive process
\begin{eqnarray*}\nonumber 
e^-(p_-) + e^+(p_+) \to H(p) + X,
\end{eqnarray*}
where $H$ is a final state particle, whose momentum $p$ is measured, but not
the spin, and $X$ is an inclusive state. 
The process is assumed to occur through an $s$-channel
exchange of a $\gamma$ and a $Z$ in the SM, and through a  new current whose 
coupling to $e^+e^-$ can be of the type $V,A$, or $S,P$, or $T$.  
We calculate the relevant factor in
the interference between the standard model currents with the 
BSM currents as
\begin{eqnarray*}\nonumber     
{\rm Tr}[(1-\g5 \hp + \g5 \sls_+)\slp_+\gamma_\mu(g_V^e-g_A^e \gamma_5)
(1+\g5 h_-+\g5 \sls_-)\slp_-\Gamma_i]H^{i\mu }.
\end{eqnarray*}
Here $g_V^e, g_A^e$ are the vector and axial-vector couplings of the
photon or $Z$ to the electron current, and $\Gamma_i$ is the
corresponding coupling to the new physics current, $p_{\pm}$ are 
the four-momenta of $e^{\pm}$, $h_{\pm}$ are the
helicities (in units of $\frac{1}{2}$) 
of $e^{\pm}$, and $s_{\pm}$ are respectively their transverse polarizations. 
For details on the notation which are spelt out in great detail,
see ref.~\cite{hep-ph/0601199}.
We should of course add the
contributions coming from photon exchange and $Z$ exchange, with the
appropriate propagator factors. However, we give here the results for
$Z$ exchange, from which the case of photon can be deduced as a special
case. The tensor $H^{i\mu }$ stands for the interference between the
couplings of the final state to the SM current and the new physics
current, summed over final-state polarizations, and over the phase space
of the unobserved particles $X$. It is only a function of the the
momenta $q=p_-+p_+$ and $p$. The implied summation over $i$
corresponds to a sum over the forms $V, A, S, P, T$, together with any
Lorentz indices that these may entail. 

We now determine the forms of the matrices $\Gamma_i$ and the 
tensors $H^{i\mu }$ in the various
cases, using only Lorentz covariance properties. 
We set the electron mass to zero.  Consider now
the three cases:

\noindent\underline {1. Scalar and Pseudoscalar case}: 
In this case, there is
no free Lorentz index for the leptonic coupling. 
Consequently, we can write it as 
\begin{eqnarray*}
\Gamma = g_S + i g_P \gamma_5.
\end{eqnarray*}
The tensor $H^{i\mu }$ for this case has only one index, viz., $\mu$.
Hence the most general form for $H$ is
\begin{eqnarray*}\nonumber
H^{S}_\mu  = F(q^2,p\cdot q) p_\mu,
\end{eqnarray*}
where F is a function of the Lorentz-invariant quantities $q^2$ and
$p\cdot q$.

\noindent\underline {2. Vector and Axial-Vector case}: 
The leptonic
coupling for this case can be written as
\begin{eqnarray*}\nonumber
\Gamma_\mu = \gamma_\mu (g_V -  g_A \gamma_5).
\end{eqnarray*}
The tensor $H$ for this case has two indices, and can be written as
\begin{eqnarray*}\nonumber
H^V_{\mu\nu} =  -g_{\mu\nu} W_1(q^2,p\cdot q) + p_\mu p_\nu
W_2(q^2,p\cdot q) + \epsilon_{\mu\nu\alpha\beta}q^{\alpha}p^\beta
W_3(q^2,p\cdot q),
\end{eqnarray*}
where now there are three invariant functions, $W_1, W_2, W_3$.

\noindent\underline {3. Tensor case}: 
In the tensor case, the leptonic coupling is 
\begin{eqnarray*}\nonumber
\Gamma_{\mu\nu} = g_T \sigma_{\mu\nu}.
\end{eqnarray*}
The tensor $H$ for this case can be written in terms of the four
invariant functions $F_1, F_2, PF_1, PF_2$ as
\begin{eqnarray*}
\begin{array}{lcl}
H^T_{\mu\rho\tau}& = & (q_\rho  p_\tau - q_\tau p_\rho ) p_\mu
F_1(q^2,p\cdot q) + ( g_{\rho\mu} p_\tau - g_{\tau\mu} p_\rho )
F_2(q^2,p\cdot q)\\&& + \epsilon_{\rho\tau\alpha\beta} p^\alpha q^\beta p_\mu
PF_1(q^2,p\cdot q) + \epsilon_{\rho\tau\mu\alpha} p^\alpha
PF_2(q^2,p\cdot q).
\end{array}
\end{eqnarray*}
Evaluating the trace in each case, we present the results in Tables
1-3, with
$\vec K\equiv (\vec{p}_- - \vec{p}_+)/2= E \hat{z}$, 
where $\hat{z}$ is a unit vector
in the z-direction, $E$ is the beam energy,
and $\vec{s}_\pm$ lie in the x-y plane.
for $g_A^e$ alone. 
The tables corresponding to $g_V^e$ alone are not given, and
was the case considered in ref.~\cite{DR} for the interference
of QED amplitudes with physics due to the then undetermined amplitude of the
neutral current due to $Z$. 

In Tables 1-3 are also given the charge conjugation C and parity P
properties of the various correlations, under the assumption that the
final-state particle observed is self-conjugate, viz., $H=\overline H$. 
If it is not
self-conjugate, then the C factor given in the tables would apply to the
sum of the cross sections for production of $H$ and $\overline H$.
The difference of these cross sections would take a C factor of the
opposite sign.
The counting of the number 
of independent correlations for the vector and axial-vector cases turns
out to be subtle, and is described at length in ref.~\cite{hep-ph/0601199}.  

\begin{table}[h!]\label{ps_gAe_table}
\begin{center}
\begin{tabular}{||c|c|c|c||}\hline
Term & Correlation & ${\rm P}$ & ${\rm C}$ \\ \hline \hline
${\rm Im}\, (g_P F)$ & $2 E^2 \,(\hp \sm+\hm \sp)\cdot \pv$ & $+$ & $+$ \\ 
${\rm Im}\, (g_S F)$ & $2 E \, [\Kv \cdot  (\hp\sm-\hm\sp)\times \pv]$ 
							& $-$ & $+$ \\
${\rm Re}\, (g_S F)$ & $2 E^2 \, \pv \cdot (\sp +  \sm)$ & $-$ & $+$ \\
${\rm Re}\, (g_P F)$ & $2 E \, [\Kv \cdot  (\sp -  \sm) \times \pv]$ & $+$ &
                            				$+$ \\ \hline
\end{tabular}
\caption{List of $S,P$ correlations for $g_{A}^e$}
\end{center}
\end{table}
\begin{table}[h!]\label{VA_gAe_table}
\begin{center}
\begin{tabular}{||c|c|c|c||}\hline
Term & Correlation & ${\rm P}$ & ${\rm C}$ \\ \hline \hline
${\rm Re}\, (g_V W_1)$ & $4 E^2(\hp - \hm ) $ & $-$ & $-$ \\
${\rm Re}\, (g_A W_1)$ & $-4 E^2(\hp \hm -1) $ & $+$ & $+$ \\
${\rm Re} \, (g_V W_2)$ & $ 2 (\Kv\cdot\Kv
         \, \pv\cdot\pv -(\pv\cdot \Kv)^2) (\hp-\hm)$
& $-$ & $-$ \\
${\rm Re} \, (g_A W_2)$ & $-2[-2 E^2 \pv\cdot \sm \pv \cdot \sp +
       (\Kv\cdot\Kv\, \pv\cdot\pv -
	(\pv\cdot\Kv)^2)(\hp\hm-1+\sp\cdot \sm)]$ & $+$ & $+$ \\
${\rm Im}\, (g_V W_3)$ & $-8 E^2(\pv \cdot \Kv)  (\hp \hm-1)$ & $+$ & $-$ \\
${\rm Im}\, (g_A W_3)$ & $ 8 E^2(\pv\cdot\Kv)  (\hp-\hm)$ & $-$ & $+$ \\
${\rm Im}\, (g_V W_2)$ & $-2E(\pv\cdot \sp [\Kv\cdot \sm\times \pv] +
                        \pv\cdot \sm [\Kv\cdot \sp\times \pv] )$ & $-$ & $-$ \\
								\hline
\end{tabular}
\caption{List of $V,A$ correlations for $g_{A}^e$}
\end{center}
\end{table}
\begin{table}[h!]\label{T_gAe_table}
\begin{center}
\begin{tabular}{||c|c|c|c||}\hline
Term & Correlation & ${\rm P}$ & ${\rm C}$ \\ \hline \hline
${\rm Im}\, (g_T F_1)$ & $-8E^2 \pv\cdot\Kv[\pv\cdot(\sp+\sm)]$ & $-$ & $-$ \\
${\rm Im}\, (g_T F_2)$ & $-4 E^2 \pv\cdot(\sp-\sm)$ & $-$ & $-$ \\
${\rm Im}\, (g_T PF_1)$ & $-8 E \pv\cdot \Kv [\Kv\cdot(\sp-\sm)\times \pv]$ &
							        $+$ & $-$ \\
${\rm Im}\, (g_T PF_2)$ & $4 E [\Kv\cdot(\sp+\sm)\times \pv]$ &
							        $+$ & $-$ \\
${\rm Re}\,  (g_T F_1)$ & $8 E \pv\cdot \Kv [ \Kv\cdot(\hp\sm-\hm\sp)\times \pv]$ &
							        $-$ & $-$ \\
${\rm Re}\,  (g_T F_2)$ & $-4 E [\Kv\cdot (\hp\sm+\hm\sp)\times \pv]$ &
							        $-$ & $-$ \\
${\rm Re}\, (g_T PF_1)$ & $8 E^2 \pv \cdot \Kv [\pv\cdot(\hp\sm+\hm\sp)]$ & $+$ & $-$ 
								\\
${\rm Re}\, (g_T PF_2)$ & $4 E^2 \pv\cdot(\hp\sm-\hm\sp)$ & $+$ & $-$ \\
								\hline
\end{tabular}
\caption{List of $T$ correlations for $g_{A}^e$}
\end{center}
\end{table}
We might like to use the behaviour of the differential cross
section to construct asymmetries which can test symmetry properties like
CP. Tables 1-3 may be employed to make some predictions for what to
expect.
It is possible to make general deductions in the special case when
the final-state is a two particle state. Within that, we consider two
possibilities:

\noindent \underline {Case 1: $H=\overline H$}
The simplest case to consider is when 
$H$ is self-conjugate, i.e., $H=\overline H$. 

\noindent \underline{Case 2: $H \neq \overline H$}
As mentioned earlier, in this case, the C properties in the tables refer
to the sum 
\begin{eqnarray*}
\Delta \sigma^+ = \Delta\sigma + \Delta\bar \sigma,
\end{eqnarray*}
where $\Delta\sigma$ and $\Delta\bar \sigma$ are partial cross sections 
corresponding respectively to $H$ and $\overline H$
production. The difference of these, 
\begin{eqnarray*}
\Delta \sigma^- =  \Delta\sigma - \Delta\bar \sigma,
\end{eqnarray*}
will have the opposite C property.

We have further considered two special cases: when the final state consists of
a pair of conjugate particles $H\overline H$, and when it consists of two
particles $H\overline H'$, where $H' \neq H$.

\noindent \underline{Case 2a: $X\equiv \overline H$}

\noindent \underline{Case 2b: $X\equiv\overline H'$, $H\neq H'$}

The specific properties are discussed in great detail in 
ref.~\cite{hep-ph/0601199}.

\section{Extensions and Applications}
So far we have dealt with a scenario where the SM interactions take
place through $s$-channel $\gamma$ and $Z$ exchanges. This is most
suitable for production of particles which have not direct coupling to
$e^-$ and $e^+$. However, for production of gauge bosons in the SM, which
couple directly to $e^+e^-$, there would be a $t$-channel and/or
$u$-channel lepton exchange.  In ref.~\cite{hep-ph/0601199} we have
provided a detailed discussion on what conclusions one may draw
regarding the correlations for this process and the CP properties
thereof, for the particular case of $Z \gamma$ production. 
The crucial factor in this
adaptation is the fact that for $m_e=0$, the only contributions which
survive correspond to opposite $e^-$ and $e^+$ helicities. For, any
final-state particles which may be emitted from an electron line with a
flip of electron helicity (as for example a Higgs boson) will have
vanishing coupling in the limit of $m_e=0$. We are thus left in the most
general case with only chirality-conserving combinations of Dirac
matrices, sandwiched between electron and positron spinors of opposite
helicities. Such a combination of Dirac matrices is a product of odd
number of them. For massless spinors, they can always be reduced to a
linear combination of $\gamma_{\mu}$ and $\gamma_{\mu} \gamma_5$. We are
thus back to the case of $V$ and $A$ couplings in the $s$
channel considered in the preceding, except that the coefficients $g_V^e$
and $g_A^e$ would now be replaced by something more complicated. In fact,
they could contain tensors constructed out of momenta occurring in the
process. It is possible to absorb these tensors into the definition of
the $H$ tensor, and 
final result would be that we could still use the tables we have
obtained so far, with appropriate redefinitions of $g_V^e$, $g_A^e$ and the
form factors.  

One further application presented in 
ref.~\cite{hep-ph/0601199} is for the 
$S,P$ and $T$ case, where it is possible to have CP-odd observables,
and of the possible ones listed therein, the ones which occur
in the special case of lowest-dimensional observables are Re$(g_P F)$
and Re$(g_T F_2)$, corresponding respectively to the four-Fermi couplings
Im$(S_{RR})$ and Im$(T_{RR})$ in the notation of \cite{hep-ph/0309260}. The
special features observed in that work, viz., that the four-Fermi
scalar coupling
terms occur with only the $g_A^e$ coupling at the electron vertex, and
that the tensor coupling terms occur with only the $g_V^e$ at the
electron vertex, are borne out by our general results.

In ref.~\cite{hep-ph/0601199}, it was further shown 
that some features of our treatment can be carried
over to an extension of SM, like the
MSSM, using as illustrations chargino
and neutralino pair production.
We have also considered
popular scenarios for BSM physics, resulting from either extra
dimensional models or from non-commutative models. 

The work reviewed here is presently being extended to 
the processes $e^+ e^- \to h_1(p_1) h_2 (p_2) X$ and
to $e^+ e^- \to h(p,s) X$.

\section{Acknowledgements:} 
We thank the Council for Scientific and
Industrial Research for support during the course of these
investigations under scheme number 03(0994)/04/EMR-II, as well
as the Department of Science and Technology, Government of India,
and K. Shivaraj and A. Upadhyay for reading the manuscript.

\end{document}